\documentclass[conference]{IEEEtran}
\usepackage{cite}
\usepackage{float}
\usepackage{amsmath,amssymb,amsfonts}
\usepackage{adjustbox}
\usepackage{times}
\usepackage{verbatim}
\usepackage{enumitem}
\usepackage{algorithmic}
\usepackage{graphicx}
\usepackage{textcomp}
\usepackage{xcolor}
\usepackage{todonotes}
\usepackage[english]{babel}
\usepackage{blindtext}
\usepackage{subfigure}
\usepackage{graphicx}
\usepackage{draftwatermark}
\SetWatermarkText{Accepted in SMC 2019}
\SetWatermarkScale{0.3}

\usepackage{multirow}
\usepackage[normalem]{ulem}
\useunder{\uline}{\ul}{}

\def\BibTeX{{\rm B\kern-.05em{\sc i\kern-.025em b}\kern-.08em
    T\kern-.1667em\lower.7ex\hbox{E}\kern-.125emX}}
\begin{document}
\title{\vspace{18pt}\LARGE \bf A Framework for Monitoring Human Physiological Response during Human Robot Collaborative Task}

\author{{Celal Savur$^{1}$ \qquad Shitij Kumar$^{2}$ \qquad Ferat Sahin$^{3}$}
\\
\IEEEauthorblockA{Department of Electrical and Microelectronic Engineering \\
Rochester Institute of Technology\\
Rochester, NY, 14623, USA \\
\{cs1323$^{1}$, spk4422$^{2}$, feseee$^{3}$\}@rit.edu }
}

\maketitle

\begin{abstract}
In this paper, a framework for monitoring human physiological response during Human-Robot Collaborative (HRC) task is presented. The framework highlights the importance of generation of event markers related to both human and robot, and also synchronization of data collected. This framework enables continuous data collection during an HRC task when changing robot movements as a form of stimuli to invoke a human physiological response. It also presents two case studies based on this framework and a data visualization tool for representation and easy analysis of the collected data during an HRC experiment.

\end{abstract}

 
\section{Introduction}
\label{sec:introduction}

The major challenges of any Human Robot Collaboration (HRC) in industry are human safety, human trust in automation, and productivity \cite{Kumar2017}. Human safety has always been the primary concern in robotics. One main aspect that concerns safety is injuries due to human-robot collision. Different strategies have been introduced to ensure human safety, one is implementing  physical and electronic safeguards according to industrial standards \cite{ISOTS15066}. However, new strategies and approaches are needed with human robot collaboration where less standards are available to implement complex protection schemes. Hence a new category of robots called \emph{collaborative robots} or \emph{cobots} have been introduced in the market (e.g. Universal Robots, Kuka lbr-iiwa, Rethink Robotics Sawyer; to name a few). These robots are purposely designed to work in direct cooperation with humans in a defined workspace by lowering the severity and risks of injury due to collision. 

Human trust in automation is about managing human expectations and how comfortable the human is sharing the robot workspace. Even though \emph{cobots} decrease the risk of injury, any form of physical collision decreases the human trust in automation. Thus, collision avoidance strategies such as stopping or reducing speeds while human is in the operating workspace of the robot have been implemented \cite{kumarDynamicAwarenessIndustrial2018} \cite{ISOTS15066}. However, the question arises how do we quantify human's trust in automation? 


 In a human robot interaction setup, change in robot  motion can affect the human behavior. This was shown in experiments done in \cite{Kulic2005} and \cite{Kulic2007a}. The literature review  in \cite{Tiberio2013} highlights the use of `psycophsiological' \footnote{Psychophysiology is a branch of neuroscience that seeks to understand how a person’s mental state and physiological responses interact to affect one another.} methods to evaluate human response and behavior during human robot interaction. In our opinion, continuous monitoring of physiological signals during human-robot task is the first step in quantifying human trust in automation. The inferences from these signals and incorporating them in real-time to affect robot motion can help in enhancing the human-robot interaction. Such a system capable of `physiological computing' \footnote{Physiological computing represents a category of `affective computing' that incorporates real-time software adaption to the psychophysiological activity of the user.} will result in a closed human-in-the-loop system where both human and robot in an HRC setup are monitored and information is shared. This could result into better communication which would improve trust in automation and increase productivity.     

Hence, in this work we propose a framework for a `physiological computing' system to monitor human physiological responses during a human-robot collaboration task. This paper highlights the aspects and challenges of collecting human-physiological signals during a human-robot experiment. It underscores the importance of a controlled HRC experiment design, event marker generation related to both human and robot, and the synchronization of data collected. In order to verify this framework, a prototype implementation of the system is shown as case studies of two HRC experiments. 

The first case study is an experiment to monitor the effect of change in robot acceleration and trajectory of motion on human physiological signals and determine a human comfort index. In this experiment the human is sitting and sharing the workspace with a UR 5e robot. The second experiment is monitoring the human-behavior for different safety algorithms during human-robot collaborative task. This task is implementation of a speed and separation monitoring setup where a human and a UR10 robot perform two separate tasks while sharing a workspace \cite{CASE2019_paper}. Here, the human is not stationary and moves in the workspace, which requires wireless data acquisition of human physiological signals and representation of human-robot shared workspace. The final objective of this work is to generate a database that can be used to further the understanding of how human physiological responses can be inferred to result in adaptive robot motion behavior.  



The remainder of the paper is organized as follows:  Section \ref{sec:approach} describes the proposed framework for creating a `physiological computing' system to monitor human physiological responses during a human-robot collaborative task. Based on this framework two case studies are implemented in Section \ref{sec:case_study} and discussed in Section \ref{sec:discussion} . Conclusions are drawn and the future work mentioned in Section \ref{sec:FutureWork}.  

\section{Proposed Approach}
\label{sec:approach}
In this section the key aspects and challenges of monitoring human physiological response in Human Robot Collaboration are presented. Asking questions to human subject during or after the experiment is common practice in human robot collaboration and interaction experiment \cite{Kulic2007, Tiberio2013}. These response of the subject allow researchers to quantify the subjective data of the experiment. However, such methods that interrupt subject during experiment may not be desirable for maintaining the integrity of the desired physiological signals. In our opinion, an alternative approach would be a system which is able to generate event markers automatically during experiment and enable the subject or the principle investigator to generate markers as the experiment is being performed. Then these event markers can be used during post processing by field expert to identify response of the given input. In this way, it could act as an alternative method to asking questions during the experiment. 

\begin{figure}[!ht]
	\centering
	\includegraphics[width=\linewidth,keepaspectratio]{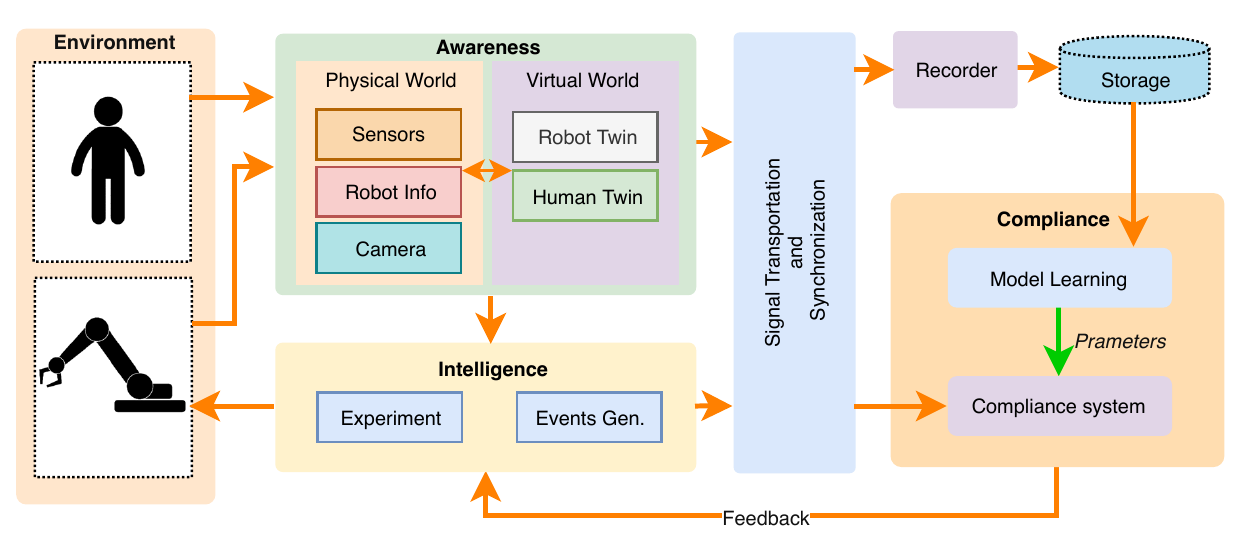}
	\caption{An overview block diagram of the proposed framework for monitoring Human Physiological Response during a Human Robot Collaborative Task.}
		\label{fig:framework}
\end{figure}

The block diagram of the proposed framework is shown in Figure \ref{fig:framework}. The proposed framework is a solution for concurrently and continuously monitoring the state of human and robot during an HRC task. The framework from a systems perspective can be conceptually categorized further into three sub modules: \textit{Awareness}, \textit{Intelligence} and \textit{Compliance} \cite{Savur2019}. The communication layer between these sub-modules is equally important as it is  responsible for data transformation and synchronization. 

The sub-module \textit{Awareness} is the perception of system which is generated from the physical world sensors and digital represented in the virtual world. The physical world is responsible to sense the environment through the sensor information such as PPG sensor, GSR sensor, camera, motion capture system etc. On the other hand virtual word is a digital-twin representation of the physical world that mimics the environment of the HRC task as well as the movements and behavior of the robot and human agents \cite{Cichon2017}. The digital twin can be used to calculate metrics such as human-robot minimum distance, directed human-robot speeds, possible collisions and changes in trajectory \cite{CASE2019_paper}\cite{safeeaMinimumDistanceCalculation2019}. The virtual world updates its state constantly based on the sensory data received from the physical world to update itself and generate new data for the framework. Overall \textit{Awareness} is responsible for sensing physical and virtual world and provide this data to rest of the system. Such a setup helps digitally represent a combined human-robot state, which can then be associated with the human physiological state.

The \textit{Intelligence} represents the control of robot actions during an HRC experiment. Programming experiment is part of the \textit{Intelligence} since it controls speed, acceleration, and trajectory of the robot. The \textit{Intelligence} module processes the  data from the \textit{Awareness} module  to generate event markers as well as robot actions that can be used as stimuli to elicit human response. In addition to \textit{Awareness} it also receives input from \textit{Compliance} module, which is a form of interpretation of human expectation. The \textit{Intelligence} module interprets this human command/feedback into actionable robot commands.

Using human physiological signal as feedback to the robot or form of actionable control will help achieve a complete human-in-the-loop closed loop system. Here, the \textit{Compliance} sub-module is responsible for inference from the physiological signals or any form of commands from the human, that can be used to modify the robot behavior. Thus achieving a higher level of \textit{Compliance} for the robot and managing the human expectation by interpreting the human physiological state can be a gateway for a more interactive human robot collaboration. 

\textit{Awareness}, \textit{Intelligence}, and \textit{Compliance} are the main parts of the framework \cite{Savur2019}, however to integrate these three modules a communication layer for data transformation and synchronization is required. This is critical as many sensor devices and other systems do not have the same frequency and timing clock. The communication layer is responsible to transfer data in real time and also synchronize the data from different sources such as physiological signal collection devices, cameras, the representation of human-robot state in the digital twin and robot state information. 

When designing human physiological signals related experiment the following aspects are critical. 
\begin{itemize}
    \item Experiment design
    \item Event markers generation
    \item Synchronization
\end{itemize}
The importance of these is elaborated in the following Sections.
\hfill

\subsubsection{Experiment design}
When designing an experiment, the experiment and its parameters need to well defined. The task need to be real or as realistic as possible to maintain the integrity of the robot motion to act as stimuli to elicit the human physiological response. For example, an industrial tasks is good option for the experiment. Hence the industrial task may improves the involvement of the subject sharing human robot collaboration workspace. In addition the task need to be simple and controlled to increase the repeatability of a human-robot interaction scenario. A complex task may result into more uncertainty.

\subsubsection{Event Marker generation}
The event markers generation is part of experiment design. In the experiment, important event need to be investigated and generated by the experiment. Having markers during experiment gives more intuition about experiment, such as \textit{“Experiment Start/End”}, \textit{“Task Start/End”}, \textit{“Robot Coming towards Human”}, etc. The event markers help to synchronize signal across different channels. For example, extracting Galvanic Skin Response and Heart Rate signal between \textit{"Experiment Start"} and \textit{"Experiment End"} is trivial when the event markers are present during signal recording. Thus, the markers can be used during post-processing for efficient data segmentation and epoching.

\subsubsection{Synchronization}
Synchronization of signal from different sensors is crucial for the human physiological response. All the signal from human and robot need to be synchronized with event markers. Thus a central synchronization system is necessary. In proposed framework for the physiological computing system, Lab Stream Layer (LSL) is used the interface the subsystems, which integrates data from all different devices being used. The Lab Stream Layer is a system for collection of time series data over a local network with built-in time synchronization \cite{SCCN2018}. The LSL stream is nearly real-time and it is commonly used in biological signal collection system such as OpenBCI, Pupil Lab, etc. Therefore, the LSL layer is selected as the central core of the data acquisition system in proposed framework. In the framework, each device has an application node that is responsible to acquire signal from the device in real-time and pushing it to the LSL stream. A node is responsible to record all-time series data from LSL stream into a local file for post-processing and analysis. Along with LSL, Robot Operating System (ROS) and ZeroMQ is used to monitor data in real time during the experiment \cite{Savur2019}. 

\section{Case Studies}
\label{sec:case_study}
\subsection{Case Study I}
\begin{figure}[ht]
	\centering
	\includegraphics[width=0.5\textwidth]{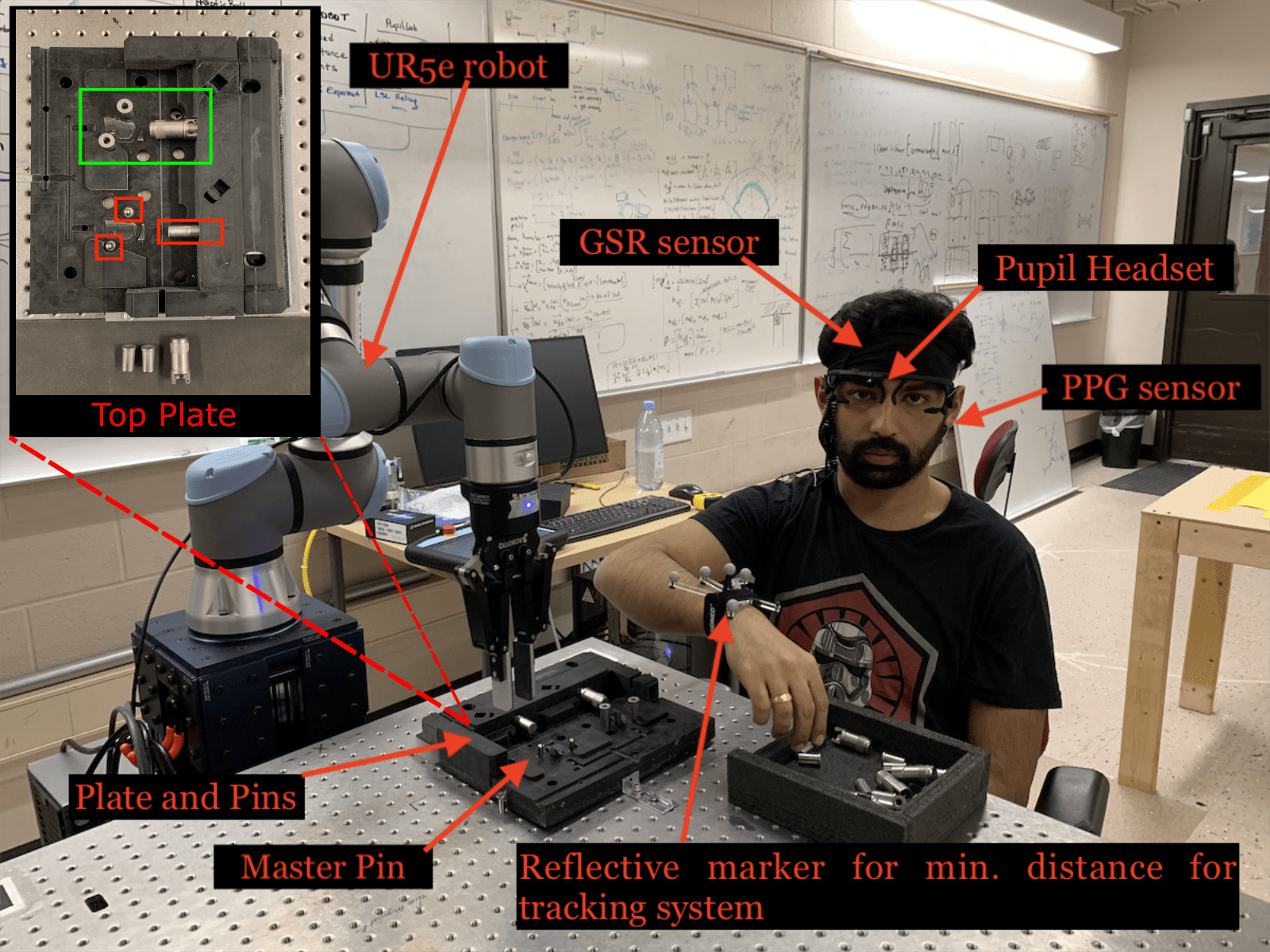}
	\caption{A picture of an subject that preparing for experiment, and device placement}
		\label{fig:real_setup}
\end{figure}

The objective of the experiment is to monitor the effect of acceleration and trajectory of the robot on human physiological signals during collaborative task. The experiment was performed using UR5e (Universal Robot) six degree of freedom (DoF) arm robot, as shown in Figure \ref{fig:real_setup}. The UR5e is a common collaborative robot with payload of 5 kg, which is suitable for manufacturing environment and laboratories. The experiment is a simplfied version of an industrial task for loading inserts and unloading parts on at a plastic injection molding plant. This experiment represents a scenario where the human robot shared workspace is on a table and the human is stationary.

The experiment consists of four sub tasks which are tabulated in table \ref{tab:tasks}. In the experiment the max speed set 100 degree/seconds so in case of collusion any injury or pain will be minimized. Since the maximum speed is fixed, the experiment is designed to have different accelerations and trajectories. In the experiment acceleration has two modes: fixed and random. The fixed mode indicates that robot has fixed acceleration and random mode means the acceleration is random. 
\begin{table}[]
\caption{The table shows the parameters of the each task}
\centering
\label{tab:tasks}
\begin{tabular}{|l|l|l|}
\hline
\textbf{}       & \textbf{Acceleration} & \textbf{Trajectory} \\ \hline
\textbf{Task 1} & Normal                & Fixed               \\ \hline
\textbf{Task 2} & High                  & Fixed               \\ \hline
\textbf{Task 3} & Normal                & Random             \\ \hline
\textbf{Task 4} & High                  & Random             \\ \hline
\end{tabular}
\end{table}

\begin{figure}[!ht]
	\centering
	\includegraphics[width=\linewidth,keepaspectratio]{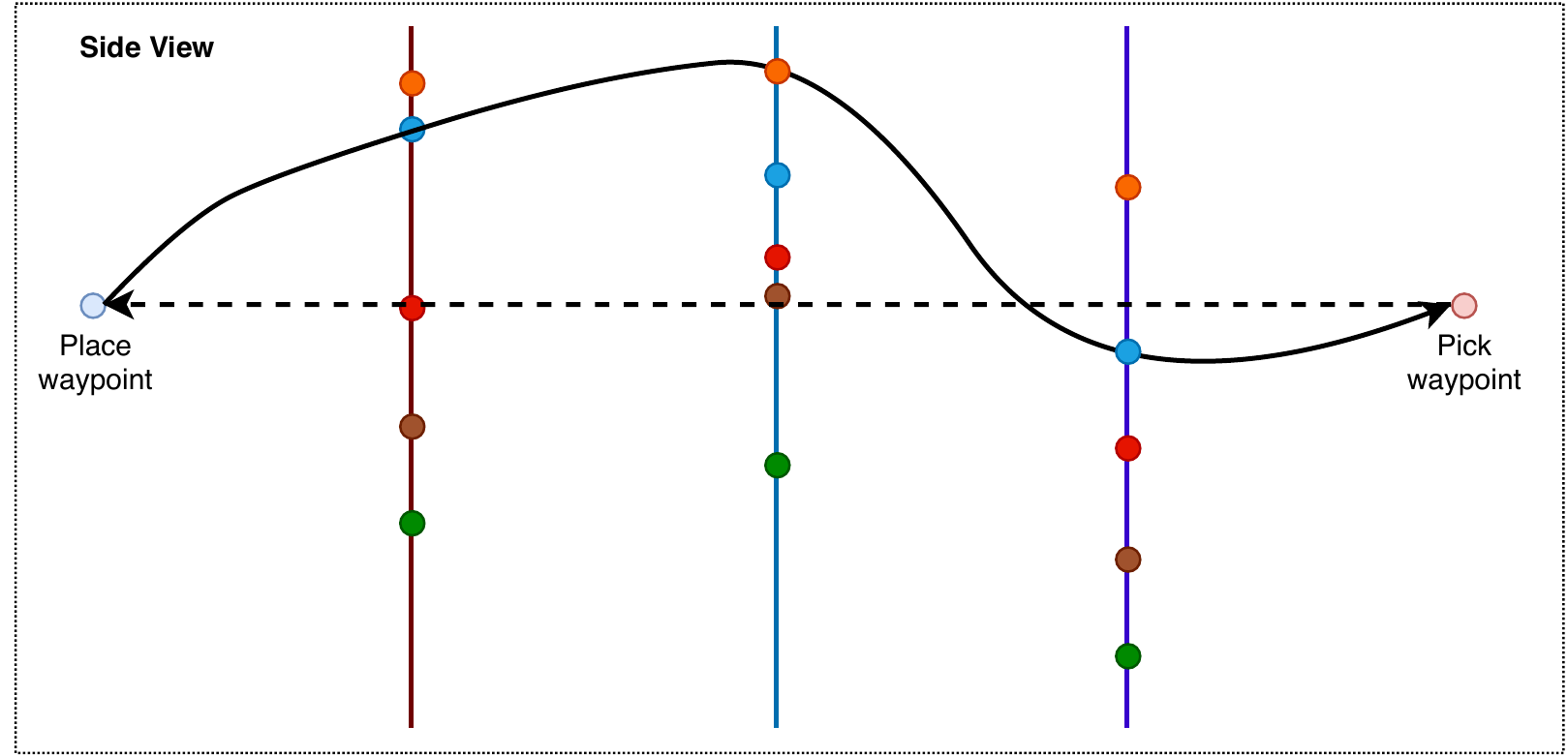}
	\caption{Figure shows how the robot selects the waypoint between pick and place positions. Solid line shows a possible random trajectory and striped line shows fixed trajectory for the robot. }
		\label{fig:trajectory_generator}
\end{figure}

The trajectory has two modes: simple and random. The simple trajectory indicates there is no waypoint between pick and place waypoints and the motion is fluent and predictable. The random trajectory indicates multiple waypoints randomly selected between pick and place waypoints. The Figure \ref{fig:trajectory_generator} shows an example of the trajectory in random mode  in which robot may take between Pick waypoint to Place waypoint or vice-versa. The trajectory planner will generate a trajectory from randomly selected waypoints from each plane.

Four type of tasks are performed by subjects. Each tasks consists of two parts: loading inserts and unloading inserts. The subject is responsible for loading inserts on plate shown in Figure \ref{fig:real_setup}(top-left). There are two possible actions which the human can take during tasks. The first one is load the plate and wait for robot to unload all of the inserts from plate then re-load the plate. The second action is, to increase the productivity, while robot is unloading inserts, load inserts that are taken by the robot. The subject has freedom to choose whichever action is comfortable.


The robot is responsible for unloading i.e. picking all inserts from the plate and placing them into the container. In order to control the start the unloading, the robot checks master pin on plate every five seconds. This is helpful in generating event-marker representing the start of the task. If the master pin has inserts then the plate is full and the robot starts the unloading process. It picks each item in order and place it into a container. If there are no inserts on master pin, the robot goes to its home position wait for five seconds. The experiment setup and sensor placement can be seen in Figure \ref{fig:real_setup}.

\subsection{Case Study II}
This experiment is monitoring the human-behavior for different safety algorithms during human-robot collaborative task. This task is implementation of a speed and separation monitoring setup where a human and a UR10 robot perform two separate but related tasks while sharing a workspace \cite{CASE2019_paper}. Here, the human is not stationary and moves in the workspace, which requires wireless acquisition of human physiological signals and representation of human-robot shared workspace.

The experiment setup is a generic robot pick and place task of placing 10 products in a box. The robot movement involves moving the base joint $180^\circ$ degrees between the pick and place positions on the tables. The human has an assembly task for threading a nut and a screw that are placed on the picking and placing area. After threading the bolts and screws the human puts the finished part on a table outside the robot workspace. This human task was setup to control the human movement and overlap of human-robot workspace. For more information our previous work \cite{kumarDynamicAwarenessIndustrial2018} and \cite{CASE2019_paper} can be referred. In order to avoid collision, safety algorithms are implemented to detect and anticipate the human motion, resulting into the robot stopping, reducing speed or moving normally i.e. maximum allowed speed for the task. The safety algorithms vary in terms of parameters such as critical human-robot separation distance and what sensors are used to calculate the separation distance. This results into different robot motion behavior.

\begin{figure}[!ht]
	\centering
	\includegraphics[width=0.4\textwidth]{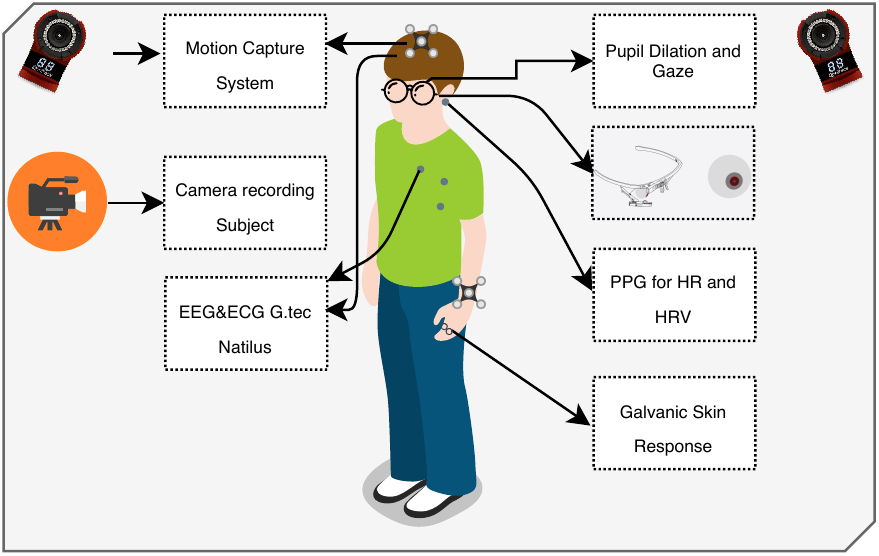}
	\caption{A motion capture is used to monitor human motion, a camera is used to record the experiment, the human-gaze is tracked using Pupil Labs and human physiological responses such as pupil dilation, PPG, GSR, EEG \& ECG recorded.}
		\label{fig:humans_signal}
\end{figure}

The objective of this experiment is to monitor human physiological response and also see the overall task productivity during this shared workspace task. During the experiment the sensors used to monitor the human are shown in Figure \ref{fig:humans_signal}. Here the motion capture is used to monitor human motion, a camera is used to record the experiment, the human-gaze is tracked using Pupil Labs and human physiological responses such as pupil dilation, PPG, GSR, EEG \& ECG recorded. 

\begin{figure*}[]
 \centering
	\includegraphics[width=0.95\textwidth,keepaspectratio]{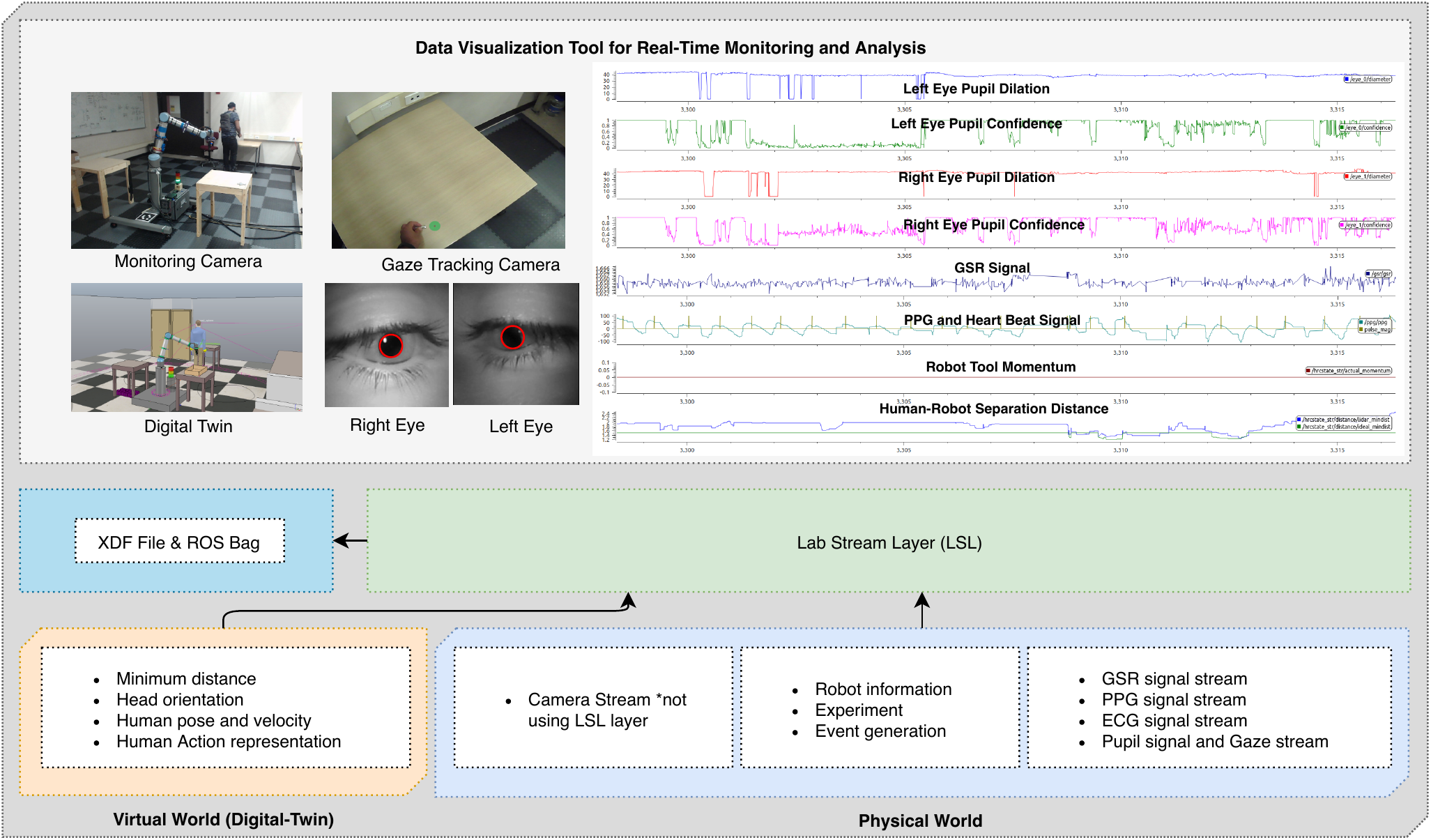}
	\caption{A system diagram representing the data collection and monitoring during the experiment as described in Case Study II}
		\label{fig:system_diagram}
\end{figure*}

A system diagram showing the data collection and monitoring is shown in Figure \ref{fig:system_diagram}. The experiment setup is represented as a digital-twin in order to represent human and robot state during the experiment. This helps in generating the human-robot interaction state data such as human-robot separation distance (minimum distance), human head orientation, human pose and velocity and action representation. This data is monitored and collected along with the human-physiological responses. It is used to represent a combined human-robot state of the `physiological computing' system and analyse the stimulus and effect of human behavior during the experiment.

In this system, the event markers used for case studies I and II, the physiological signals that can be used and the communication and synchronization of data are discussed in the following section.


\section{Discussion}
\label{sec:discussion}
\subsection{Event Marker}
The auto generation of event markers during an HRC experiment is critical. The choice of event markers depends on the experiment setup and the objective of the experiment. The biggest advantage of auto generation of event markers is the experiment can be performed uninterrupted. These event markers can be used to effectively post-process and analyse the data as data segmentation and epoching of the collected signals becomes easier.
A list of events that are automatically generated during the HRC task for Case Studies I and II are listed in Table \ref{tab:event_markers}.

\begin{table}[]
\caption{The table shows the event-markers used in Case Study I \& II}
\centering
\resizebox{\linewidth}{!}{%
\begin{tabular}{|c|l|l|}
\hline
\textbf{}                     & \multicolumn{1}{c|}{\textbf{Event Marker}} & \multicolumn{1}{c|}{\textbf{Definition}}                                                                 \\ \hline
\multirow{8}{*}{ \rotatebox[origin=c]{90}{\textbf{Case Study I}}}  & Experiment start                           & Experiment started                                                                                       \\ \cline{2-3} 
                                        & Task {[}n{]} init                          & \begin{tabular}[c]{@{}l@{}}nth task initialized but subject has not \\ complete loading yet\end{tabular} \\ \cline{2-3} 
                                        & Task {[}n{]} start                         & nth task started robot unloading all the parts                                                           \\ \cline{2-3} 
                                        & Task {[}n{]} end                           & nth task unloading is done                                                                               \\ \cline{2-3} 
                                        & Robot approaching                          & \begin{tabular}[c]{@{}l@{}}Each time robot comes toward human will \\ generate a event\end{tabular}      \\ \cline{2-3} 
                                        & Pick up successful                         & Master pin is loaded                                                                                     \\ \cline{2-3} 
                                        & Pick up failed                             & Master pin is not loaded                                                                                 \\ \cline{2-3} 
                                        & Experiment end                             & Experiment is complete                                                                                   \\ \hline
\multirow{6}{*}{\rotatebox[origin=c]{90}{\textbf{Case Study II}}} & Experiment start                           & Experiment                                                                                               \\ \cline{2-3} 
                                        & Robot state change                         & \begin{tabular}[c]{@{}l@{}}When robot change state between Normal, \\ Reduced, and Stop\end{tabular}     \\ \cline{2-3} 
                                        & Robot is stopping                          & When robot going to complete stop                                                                        \\ \cline{2-3} 
                                        & Robot is speeding up                       & When robot is going to normal speed                                                                      \\ \cline{2-3} 
                                        & Robot is slowing down                      & When robot is slowing down.                                                                              \\ \cline{2-3} 
                                        & Experiment end                             & Experiment is complete                                                                                   \\ \hline
\end{tabular}
}
\label{tab:event_markers}
\end{table}

\subsection{Physiological Signals}
\label{sec:physiological_signals}
In this Section, we list some of the human physiological signals that have been used during human-robot experiments. The devices for collecting these signals have been successfully interfaced in the implemented prototype system of the proposed framework.

\begin{list}{\textbullet}{\leftmargin=0.5em}
	\item \textbf{Electroencephalogram (EEG)} is the method to record the brain's electrical activity via non-invasive electrode placed on the human head. EEG has been used for error related potentials, emotional valence scale and evoked potentials. It has also been used to detect alpha activity, which determines attentiveness, stress, and other emotions. It can be questioned that wearing an EEG cap while working can be uncomfortable. However, it must be noted that in industry, workers can wear helmets or hats. With the advent of advance, IoT systems wireless communication and small size factor of EEG equipment make it plausible to get such data. e.g., g.Tec, BioRadio, and openBCI. 
	\item \textbf{Electrocardiogram (ECG)} measures the heart's electrical activity. ECG can be used as a psychophysiological indicator for physical stress, mental stress and fatigue. In an industrial setup, robot behavior can be adjusted based on the state of health of the operator. This can help in avoiding injuries that may result from work exhaustion. \cite{Ali2018}. 
	\item \textbf{Electromyography (EMG)} is method to record electrical activity generated by muscles. EMG have been used as a control input for basic robot interaction. A sense of control is very important for building the trust of human. Another example of EMG is using facial muscles to give information about sudden emotional change or reaction.  Placement of these can be in safety glasses worn by the operator \cite{Kulic2007, Gouizi2011, Savur2017} \cite{Chalapathy2019}.
	\item \textbf{Galvanic Skin Response (GSR)} also known as Skin Conductivity (SC) or Electro Dermal Activity (EDA), measures skin conductivity which is triggered by the central nervous system. This signal has been used in for emotion recognition, lie detector and detecting physical and mental stress \cite{Kulic2007,Rohrmann1999,Kim2004,Ali2018} \cite{VanDooren2012}.
	\item \textbf{Heart Rate (HR) and Heart Rate Variability (HRV)} is a signal that can be extracted from the ECG and also photoplethysmogram (PPG) signal. This information can give the state of the person i.e. Resting or Active. HRV has been used as a psychophysiological indicator.
	\item \textbf{Pupil Dilation} is a measurement of pupil diameter change. The pupil dilation can be caused by ambient light change in environment and emotional change. \cite{Bonifacci2015}.
\end{list}

\subsection{Data Transfer and Signal Synchronization}
The proposed framework in Figure \ref{fig:framework} for monitoring human response during Human Robot Collaborative task uses LSL layer as the core for transportation and synchronization. Using LSL layer as the core brings many advantages. The first and most important reason is that it has built in time synchronization. In addition to synchronization, it allows developer to use external timer as well. The second most important feature is the LSL layer is operating system agnostic. This bring flexibility to the proposed framework, since there are sensor manufacturers have device drivers that supports only certain operating systems.

Although LSL layer has the ability to record signal from the stream as an XDF file, the proposed framework uses ROSbag as an alternative for recording. Rosbag is a popular tool in robotic application to record time-series data and replaying data from collected bags. In addition, it has tools helps plotting the stream from the bags. Hence it is selected as parallel recording with LSL layer.

Figure \ref{fig:system_diagram} shows proposed framework. In the Figure each device has an application node which push data to LSL layer. Then LSL layer deliver data to two receivers, LabRecorder and LSL2Bag application which are responsible to record data in to a file. 


\section{Conclusion and  Future work}
\label{sec:FutureWork}
In this research, a framework for monitoring and collecting human physiological response during human robot collaborative task is presented and a prototype implementation is shown. The challenges of data communication, signal synchronization and event markers are addressed and solution proposed. The implementation shows the synchronized and continuous collection of human-robot states and human physiological responses. This system is expandable for additional sensors. Although the framework designed for human robot collaboration task, it is not limited to this setup. Similar approach can be taken for other `physiological computing' systems.

Future research will focus on developing a complete user interface application of the `physiological computing' system for processing of recording signals, extracting information and applying machine-learning algorithm to provide feedback to the robot. The final objective of this work is to generate a database that can be used to further the understanding of how human physiological responses can be inferred to result in adaptive robot motion behavior.

\section*{Acknowledgment}
The authors would like to thank the Electrical Engineering Department at RIT. The authors are grateful to the staff of Multi Agent Bio-Robotics Laboratory (MABL) and the CM Collaborative Robotics Research (CMCR) Lab for their valuable inputs. 

\bibliographystyle{IEEEtran}
\bibliography{CelalsBibtex,ShitijsBibtex1,ShitijsBibtex2,papers_v1}{}
\bibliographystyle{plain}

\end{document}